\documentclass[superscriptaddress,runinaddress,showpacs,showkeys,aps,prd,floatfix,onecolumn]{revtex4}
\usepackage{amssymb,amsmath,epsfig}
\usepackage{amsmath,graphicx}
\usepackage{changes,palatino}

\usepackage[colorlinks=true,
            linkcolor=blue,
            urlcolor=blue,
            citecolor=blue]{hyperref}
\begin{document}

\title{Weak Gravitational lensing by stringy black holes}

\author{Wajiha Javed}
\email{wajiha.javed@ue.edu.pk; wajihajaved84@yahoo.com} 
\affiliation{Division of Science and Technology, University of Education, Township-Lahore, Pakistan}

\author{Muhammad Bilal Khadim}
\email{blaljutt723@gmail.com}
\affiliation{Division of Science and Technology, University of Education, Township-Lahore, Pakistan.}

\author{Jameela Abbas}
\email{jameelaabbas30@gmail.com}
\affiliation{Department of Mathematics, University of Education,\\
Township, Lahore-54590, Pakistan.}

\author{Ali {\"O}vg{\"u}n}
\email{ali.ovgun@pucv.cl}
\affiliation{Instituto de F{\'i}sica, Pontificia Universidad Cat{\'o}lica de
Valpara{\'i}so, Casilla 4950, Valpara{\'i}so, Chile.}
\affiliation{Physics Department, Arts and Sciences Faculty, Eastern Mediterranean
University, Famagusta, North Cyprus via Mersin 10, Turkey.}

\date{\today}

\begin{abstract}
In this paper, we discuss the weak gravitational lensing in the
context of stringy black holes. Initially, we examine the deflection angle of photon by charged
stringy black hole. For this desire, we compute the Gaussian
optical curvature and implement the Gauss-Bonnet theorem to investigate
the deflection angle for spherically balanced spacetime of stringy black hole.
We also analyze the influence of
plasma medium in the weak gravitational lensing for stringy black hole. Moreover,
the graphical impact of impact parameter $b$ ,
black hole charge $Q$ on deflection angle by charged stringy black hole has been
studied in plasma as well as non-plasma medium.
\end{abstract}


\maketitle

\section{Introduction}
Einstein in $1916$ announced about the presence of the gravitational lensing as well as gravitational
waves as a part of the theory of general relativity (GR).
His theory depends on many experimental and observational tests.
In accordance with his theory, the force of gravity due to the fact of curvature
in spacetime and gravitational waves \cite{1} are ripples
in the fabric of a universe which are obtained by the clashing
of the gigantic objects such like black holes (BHs). The phenomenon in which gravity deflects
light is referred as \textit{gravitational lensing}. Gravitational lensing is very
useful method to understand dark matter, galaxies, and universe. After the first observation
of Eddington, so many works have been done on gravitational lensing for black holes, wormholes
and other objects. The topic of gravitational lensing take a great attraction. Therefore,
the gravitational lensing is categories into two ways; strong gravitational lensing and
other one is a weak gravitational lensing. The strong gravitational lensing tells us about
the magnification, position and time delays of the images by BHs. Strong lensing phenomena
also need in many cases which provide further information from experimental frame of reference
to see the other different objects such as monopoles \cite{2}, boson stars \cite{3,4,5}, fermion stars \cite{6,7}
etc. On the other hand the weak gravitational lensing give a method to find the mass of astronomical
objects without demanding to their formation or dynamical states. Weak lensing also distinguish between
dark energy and modified gravity and also examine the cause of rapid expansion in the universe.
The gravitational lensing is by means of the distribution of
matter among the source and observer, that is qualified
by deflection of light from the source to the observer
also the deflection of light is one of the prediction of Einstein`s
theory \cite{Badia:2017art,Latimer:2013rja,Elghozi:2016wzb,Ahmedov:2019dja,Turimov:2018ttf,Abdujabbarov:2017pfw,Schee:2017hof,Ghaffarnejad:2016dlw,Aazami:2011tw,Virbhadra:2007kw,Keeton:2006sa,Keeton:2005jd,Bhadra:2003zs,Cao:2018lrd,Lim:2016lqv,Sultana:2013fda,Fleury:2017owg,Whisker:2004gq,Chen:2009eu,Eiroa:2002mk,Wang:2019cuf,Mao:1991nt,Bozza:2002zj,Sharif:2015qfa,Virbhadra:2002ju,Kasikci:2018mtg,Zhang:2018yzr}. The idea first given by German astronomer Johan Georg Van Soldner $1801$ in the background of Newtonian theory.
In $1924$ Chwolson and Einstein in $1936$ derived the exact calculations of deflection of light in the framework of GR.
Regarding to this, Gibbons and werner have investigated the deflection angle of light
for asymptotically flat static BHs \cite{8} by using Gauss-Bonnet theorem which is defined as:
\begin{equation}
\hat{\alpha}= -\int\int_{D_{\infty}}\mathcal{K} d\tilde{\sigma},\nonumber
\end{equation}
where $\mathcal{K}$ stand for Gaussian curvature and $d\tilde\sigma$ is a surface element of optical metric.
After that, Werner expanded this strategy for stationary BHs \cite{Werner:2012rc}.
Arakida and Kasai  \cite{10} have analyzed the engagement of cosmological constant
on deflection of light also its status with in the cosmological lens equation.
In recent times, Bisnovatyi-Kogan and Tsupko \cite{11} have showed that the deflection angle in the
homogeneous plasma is differ from the deflection angle in a vacuum, in addition that the deflection
angle in plasma medium found on the photon frequency. Additionally,
this procedure has been prolonged to the wormhole geometries
and non-asymptotically flat spacetimes having topological defects \cite{wh1,wh2,wh3,wh4,wh6,wh7,wh8}.\\
According to GR, spacetime singularities rise various issues, both
scientific and physical \cite{14,15}, by applying the nonlinear electrodynamics
it is reasonable to solve these singularities by fabricating a regular
BH solution \cite{16,17,177}. He and Lin \cite{He:2016vxc} have investigated
the deflection of light for test particles, due to a axially moving
Kerr-Newman BH with an arbitrary constant velocity that is perpendicular to its
angular momentum. Additionally, it is demonstrated that here is no peculiarity
of the electric field quality at the cause for the point-like particles and it
has an attractive charge. By using the Gauss-bonnet theorem
GBT, deflection angle of light on the foundation of magnetized BH and impact of non-linear
electrodynamic has been found by Javed et al.\cite{19,199}. Recently, plasma medium
deflects photons shown by Crisnejo and Gallo \cite{20}. By investigating
the weak gravitational lensing for hairy BH in the back burner of Einstein-Maxwell
theory (EMT) with a non-minimally coupled dilaton and its non-trivial potential, in this connection
by virtue of GBT deflection angle of light has been computed with
plasma and excluding plasma medium by Javed et al \cite{1999}. For more new works, one can see \cite{Ishihara:2016vdc,Ishihara:2016sfv,Arakida:2017hrm,J1 20,J1 21,Ovgun:2019wej,Ono:2017pie,Jusufi:2017vta,Ono:2018jrv,Jusufi:2017vew,Ono:2018ybw,Jusufi:2017lsl,Li:2019mqw,Jusufi:2017hed,Li:2019vhp,Jusufi:2018jof,deLeon:2019qnp,plasma,Jusufi:2017uhh,Crisnejo:2019xtp,Ovgun:2018prw,Ovgun:2018fte,Ovgun:2018tua,Javed:2019a,Javed:2019jag,Javed:2019rrg,Kumaran:2019qqp,km2,km3,Ovgun:2019qzc}.
Main cause of this work is to study the stringy black holes by utilizing Gauss-Bonnet theorem
and also to see the effect of topological defects on gravitational lensing.
For examination, we consider the notation of the deflection angle of big particles and
the deflection of photons in a plasma medium.\\
This paper arranged as: in section 2, We concisely review regarding stringy black holes.
In section 3, we apply Gauss-Bonnet theorem to find deflection angle of stringy black holes.
In section 4, we enhance our work to find out the deflection angle in the presence of plasma medium
and in addition we demonstrate the graphical impact of deflection angle in the context of stringy black hole.
In section 5, we recap our results which we obtain in present work.
\section{Computation Of weak lensing by Stringy black holes and Gauss-Bonnet theorem}
The stringy black holes metric in spherically coordinate is given as \cite{25};
\begin{equation}
ds^{2}=\frac{dt^{2}}{F(r)^2}-F(r)^{2}\{dr^{2}+r^{2}(d\theta^{2}+\sin^{2}\theta d\phi^{2})\},
\end{equation}
where $F(r)$ is
\begin{equation}
F(r)=1+\frac{Q}{r}+\frac{Q^2\alpha}{8r(r+Q)^3}.\\
\end{equation}
Note that $Q$, and $\alpha$ is charge of Stringy black hole and coupling constant respectively.
The optical metric is simply written in equatorial plane ($\theta=\frac{\pi}{2}$) to get null geodesics ($ds^2=0$)
\begin{equation}
    dt^{2}=F(r)^4dr^2+r^{2}F(r)^{4}d\phi^2.
\end{equation}

The Gaussian optical curvature that is evaluated as follows:
\begin{equation}
          \mathcal{K}=\frac{RicciScalar}{2}.\\
\end{equation}
After simplifying, Gaussian optical curvature is stated as:
\begin{equation}
    \mathcal{K}\approx -2\,{\frac {Q}{{r}^{3}}}+ \left( 12\,{r}^{-4}-4\,{\frac {\alpha}{{r}^{
6}}} \right) {Q}^{2}
+\mathcal{O}(Q^3,\alpha^2,r^{7}).\label{AH6}
\end{equation}

Now, we derive the deflection angle of a stringy BH by using the GBT.
By applying the GBT to the region $\mathcal{N}_{R}$, given as \cite{8}
\begin{equation}
\int_{\mathcal{N}_{R}}\mathcal{K}dS+\oint_{\partial\mathcal{N}_{R}}kdt
+\sum_{k}\sigma_{k}=2\pi\mathcal{X}(\mathcal{N}_{R}),
\end{equation}
where, $k$ is geodesic curvature, $\mathcal{K}$ represent Gaussian curvature respectively, and $k$ is defined as
$k=\bar{g}(\nabla_{\tilde{\alpha}}\tilde{\alpha},\bar{\alpha})$ in such a way that the Riemannian metric $\bar{g}
(\tilde{\alpha},\tilde{\alpha})=1$, here $\bar{\alpha}$ is unit acceleration vector and $\sigma_{k}$ represent the exterior angle
of $k^{th}$ vertex respectively. When $R\rightarrow\infty$ the jump angles equal to
$\pi/2$, hence $\sigma_{O}+\sigma_{S}\rightarrow\pi$. Where $\mathcal{X}(\mathcal{N}_{R})=1$ is a Euler
characteristic and $\mathcal{N}_{R}$ represents the non singular. Thus we obtian
\begin{equation}
\int\int_{\mathcal{N}_{R}}\mathcal{K}dS+\oint_{\partial
\mathcal{N}_{R}}kdt+\sigma_{k}=2\pi\mathcal{X}(\mathcal{N}_{R}).
\end{equation}
Here, $\alpha_{\bar{g}}$ is a geodesic and the total jump angle is $\sigma_{k}=\pi$,
since $\mathcal{X}$ represent  Euler
characteristic number which is $1$. When
$R\rightarrow\infty$ then remaining part is
$k(D_{R})=\mid\nabla_{\dot{D}_{R}}\dot{D}_{R}\mid$.
Considering the radial component for geodesic curvature that is described as:
\begin{equation}
(\nabla_{\dot{D}_{R}}\dot{D}_{R})^{r}=\dot{D}^{\phi}_{R}
\partial_{\phi}\dot{D}^{r}_{R}+\Gamma^{r}_{\phi\phi}(\dot{D}^{\phi}_{R})^{2}.
\end{equation}
At very high $R$, $D_{R}:=r(\phi)=R=const$. Thus, the leading
term of above equation vanishes and $(\dot{D}^{\phi}_{R})^{2}
=\frac{1}{f^2(r^\star)}$. Recalling $\Gamma^{r}_{\phi\phi}=
    \frac{-2r^2F^\prime}{F}-r$, we get
\begin{equation}
(\nabla_{\dot{D}^{r}_{R}}\dot{D}^{r}_{R})^{r}\rightarrow\frac{1}{R}.
\end{equation}
And which proves that the geodesic curvature is not effected to the topological
defects (i.e $k(D_{R})\rightarrow R^{-1}$). We can write $dt=Rd\phi$. Thus;
\begin{equation}
k(D_{R})dt=\frac{1}{R}Rd\phi.
\end{equation}
From the pervious results, we get
\begin{equation}
\int\int_{\mathcal{N}_{R}}\mathcal{K}ds+\oint_{\partial\mathcal{N}_{R}}kdt
=^{R \rightarrow\infty }\int\int_{S_{\infty}}\mathcal{K}dS+\int^{\pi+\sigma}_{0}d\phi.\label{hamza2}
\end{equation}
In the weak field deflection limit at the zeroth order
the light ray is described as $r(t)=b/\sin\phi$.
Therefore, the deflection angle stated as: \cite{8}
\begin{equation}
\tilde{\alpha}=-\int^{\pi}_{0}\int^{\infty}_{b/\sin\phi}\mathcal{K}\sqrt{det\bar{g}}dud\phi.\label{AH7}
\end{equation}

We substitute the leading term of equation \ref{AH6} into above equation \ref{AH7},
so the obtained deflection angle up to leading order term is stated as:
\begin{equation}
\tilde{\alpha}= {\frac {4Q}{b}}-\,{\frac {3{Q}^{2}\pi}{{b}^{2}}}
+\mathcal{O}(Q^3,b^3). \label{13}
\end{equation}

\section{Graphical Influence of deflection angle upon stringy black hole}
In this portion, we analyze the graphical behavior of deflection
angle $\tilde{\alpha}$ on stringy BH. We also give the physical importance of these
graphs to examine the effect of impact parameter $b$, and black holes charge $Q$ on deflection angle.
\subsection{Deflection angle $\tilde{\alpha}$ w.r.t impact parameter $b$}

\begin{figure}
    \centering
    \includegraphics{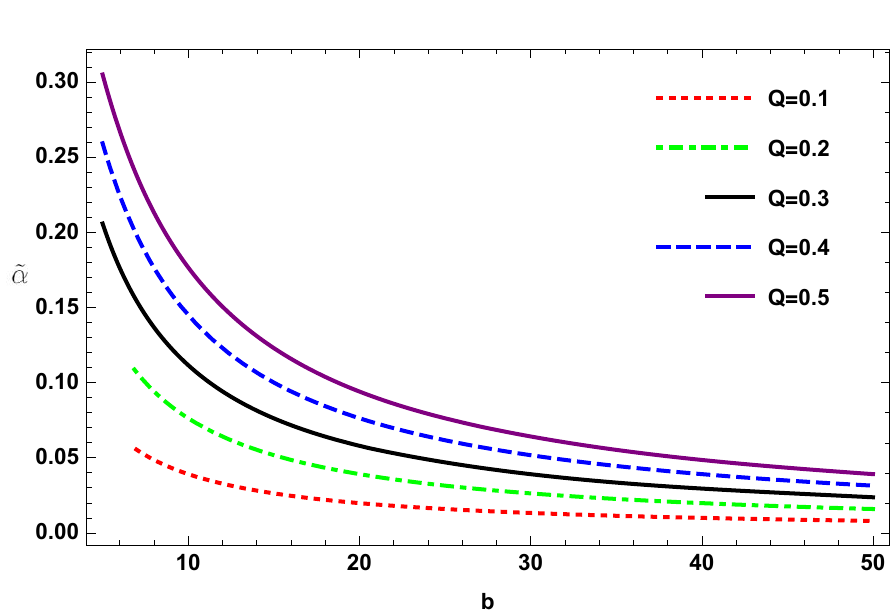}
        \includegraphics{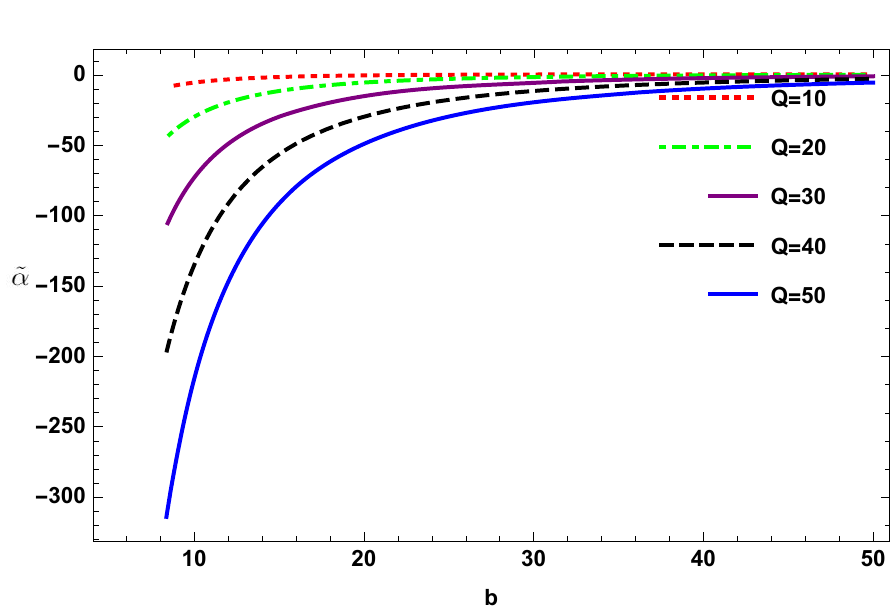}
    \caption{Figure 1: $\tilde{\alpha}$ versus $b$}.
    \label{fig:1}
\end{figure}

\begin{itemize}
\item  \textbf{Figure 1}, the left plot shows the behavior of
deflection angle $\sigma$ w.r.t $b$ for fixed BH charge $Q$ has been checked.\\
The right plot renders the behavior
of deflection angle $\tilde{\alpha}$ with $b$ by taking varying $Q$. We analyzed that the deflection
angle exponentially decreasing and approached to positive
infinity. Thus the deflection angle get the stable behavior for positive values.
\end{itemize}
\section{Weak Lensing By Stringy Black Holes In a Plasma Medium}
This segment is based on the investigation of
weak gravitational lensing of Stringy black hole in the presence of plasma medium. For
Stringy black holes the refractive index $n(r)$, is obtain as, \cite{plasma}
\begin{equation}
n(r)=\sqrt{1-\frac{\omega_{e}^{2}}{\omega_{\infty}^{2}}F(r)^{-2}},
\end{equation}
where $\omega_{e}$ and $\omega_{\infty}$ are electron plasma
frequency and photon frequency measured by an observer at
infinity respectively, then the corresponding optical metric illustrated as
\begin{equation}
dt^{2}=g^{opt}_{ij}dx^{i}dx^{j}=n^{2}(r)[F(r)^{4}dr^{2}+r^{2}F(r)^{4}d\phi^{2}].
\end{equation}
Here the metric function of the above optical metric is given as
\begin{equation}
F(r)=1+\frac{Q}{r}+\frac{Q^2\alpha}{8r(r+Q)^3}.\\
\end{equation}

Now, the value of Gaussian curvature is found as:
\begin{eqnarray}
\mathcal{K}&\approx& -2\,{\frac {Q}{{r}^{3}}}+ \left( 12\,{r}^{-4}-4\,{\frac {\alpha}{{r}^{
6}}} \right) {Q}^{2}+
\frac{\omega_e^2}{\omega_\infty^2}\left(\frac{-3Q}{2r^3}+\frac{8Q^2}{r^4}
-\frac{3Q^2\alpha}{r^6}\right)+\mathcal{O}(Q^3,\alpha^2).
\end{eqnarray}

For this, we use GBT to compute the deflection
angle in order to relate it with non-plasma. As follows, for calculating
angle in the weak field area, as the light beams become a straight line
. Therefore, we use condition of $ r=\frac{b}{sin\phi}$ at zero order.
\begin{equation}
\tilde{\alpha}=-\lim_{R\rightarrow 0}\int_{0} ^{\pi} \int_\frac{b}{\sin\phi} ^{R} \mathcal{K} dS
\end{equation}
So, the deflection angle in the presence of plasma medium is defined as;
\begin{eqnarray} \label{19}
\tilde{\alpha}=& 
\frac {4Q}{b}-\,{\frac{3{Q}^{2}\pi}{{b}^{2}}}
+\frac{3Q}{b}\frac{\omega_{e}^{2}}{\omega_{\infty}^{2}}
-\frac{2Q^2}{b^2}\frac{\omega_{e}^{2}}{\omega_{\infty}^{2}}+\mathcal{O}(Q^3,b^3).
\end{eqnarray}

The above results tells us that photon rays are moving into medium of
homogeneous plasma. We see that equation \ref{19} reduced into equation
\ref{13} when plasma effect is removed.

\section{Graphical Analysis for plasma medium}
This section give us detailed graphical analysis of deflection angle $\tilde{\alpha}$
in the presence of plasma medium. We discussed the effect of
black holes charge $Q$, impact parameter $b$ and $\beta$ on deflection angle 
for this, we take $\beta=\frac{\omega_{e}^{2}}{\omega_{\infty}^{2}} =10^{-1}$
\begin{figure}
    \centering
    \includegraphics{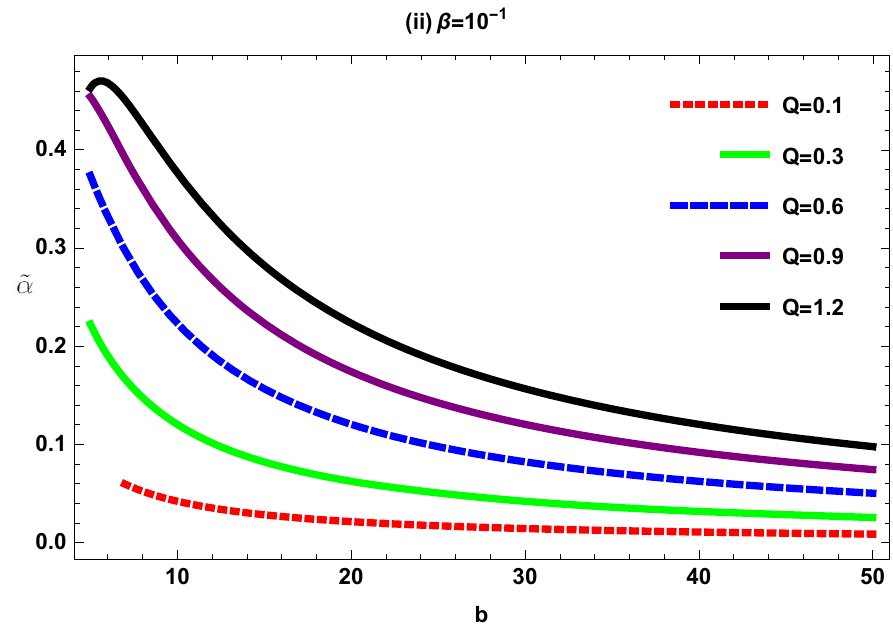}
        \includegraphics{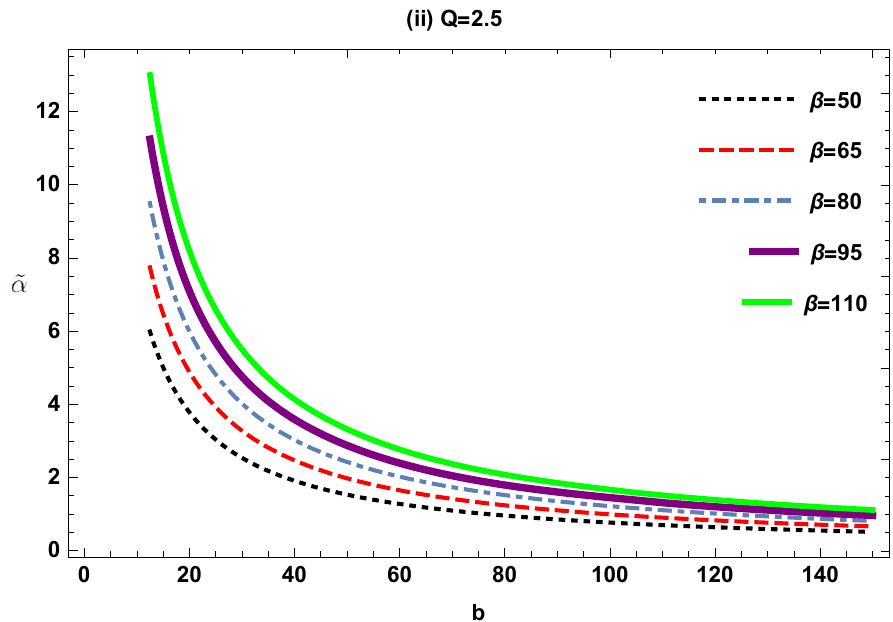}
    \caption{Figure 2: $\tilde{\alpha}$ versus $b$}.
    \label{fig:2}
\end{figure}

\begin{itemize}
\item  \textbf{Figure 2}, the left plot provide us behavior of deflection angle $\tilde{\alpha}$ with
impact parameter $b$ by changing value of $Q$ and taking fixed value of $\beta$.
We determine that for small values of $Q$ the deflection angle exponentially decreasing
and then approached to positive infinity, but as we choose high values of $Q$
the behavior of deflection angle is unstable. Thus, we summarized that as we
choose small values for $Q$ the behavior of deflection is stable. 
The behaviors of deflection
angle $\tilde{\alpha}$ with $b$ by changing the values of $\beta$ and for fixed $Q$ has been analyzed in right plot. We
examined that the deflection angle decreasing gradually and move towards positive infinity.

\end{itemize}

\section{Conclusion}
The present paper is about the investigation of deflection
angle by stringy BH in plasma and non-plasma medium.
Initially, we study the weak gravitational lensing by using
GBT and derived the deflection angle of photon. Then,
our proposed deflection angle is as follows,
\begin{equation}
\tilde{\alpha}= {\frac {4Q}{b}}-\,{\frac {3{Q}^{2}\pi}{{b}^{2}}}
+\mathcal{O}(Q^3,b^3).\nonumber 
\end{equation}
The above deflection angle is reduced to Schwarzschild-like deflection angle respect to first-order terms when $Q=M$. We have also discussed the graphical effect
of different parameters on deflection angle by stringy BH. We have observed that, there is a direct relation between deflection angle and impact parameter
while to obtain the stable behavior of deflection angle we only choose $0<Q\leq1$ and negative $\alpha$.\\
We have managed to expand the range of our research by comprising the influence of the plasma medium also on the angle of deflection given by Eq. $(19)$ which is;
\begin{eqnarray}
\tilde{\alpha}=& 
{\frac {4Q}{b}}-\,{\frac {3{Q}^{2}\pi}{{b}^{2}}}
+\frac{3Q}{b}\frac{\omega_{e}^{2}}{\omega_{\infty}^{2}}
-\frac{2Q^2}{b^2}\frac{\omega_{e}^{2}}{\omega_{\infty}^{2}}+\mathcal{O}(Q^3, b^3).\nonumber
\end{eqnarray}
If we neglected the $\frac{\omega_{e}}{\omega_{\infty}}$ term plasma effect can be removed.
Additionally, we have demonstrated the graphical impact of deflection
angle on stringy BH in a plasma medium.

\acknowledgments
This work was supported by Comisi{\'o}n Nacional de Ciencias y Tecnolog{\'i}a of Chile through FONDECYT Grant $N^\mathrm{o}$ 3170035 (A. {\"O}.).

\end{document}